\documentclass[10pt, conference]{IEEEtran}

\usepackage{graphicx}
\usepackage{color}
\usepackage{float}

\usepackage{url}
\usepackage{comment}
\usepackage{hyperref}

\usepackage[multiple,hang,flushmargin]{footmisc} 

\usepackage{eurosym}

\makeatletter
\renewcommand\footnoterule{%
  \kern-3\p@
  \hrule\@width.4\columnwidth
  \kern2.6\p@}
  \makeatother

\usepackage{balance} 

\title{Privacy Issues and Data Protection in Big Data: A Case Study Analysis under GDPR}

\author{\IEEEauthorblockN{Nils Gruschka\IEEEauthorrefmark{2}, Vasileios Mavroeidis\IEEEauthorrefmark{2}, Kamer Vishi\IEEEauthorrefmark{2}, Meiko Jensen\IEEEauthorrefmark{1}} \IEEEauthorblockA{\IEEEauthorrefmark{2}Research Group of Information and Cyber Security, University of Oslo, Norway\\ Email(s): \{nilsgrus, vasileim, kamerv\}@ifi.uio.no}
\IEEEauthorblockA{\IEEEauthorrefmark{1}Faculty of Computer Science and Electrical Engineering, Kiel University of Applied Science, Germany
\\Email: meiko.jensen@fh-kiel.de}
}

\begin{document}

\maketitle
\thispagestyle{plain}
\pagestyle{plain}

\begin{abstract}
Big data has become a great asset for many organizations, promising improved operations and new business opportunities. However, big data has increased access to sensitive information that when processed can directly jeopardize the privacy of individuals and violate data protection laws. As a consequence, data controllers and data processors may be imposed tough penalties for non-compliance that can result even to bankruptcy. 
In this paper, we discuss the current state of the legal regulations and analyse different data protection and privacy-preserving techniques in the context of big data analysis. In addition, we present and analyse two real-life research projects as case studies dealing with sensitive data and actions for complying with the data regulation laws. We show which types of information might become a privacy risk, the employed privacy-preserving techniques in accordance with the legal requirements, and the influence of these techniques on the data processing phase and the research results. 
\end{abstract}

\vspace*{.2em} 

\section{Introduction}
\label{sec:intro}

The term big data describes large or complex volumes of data, both structured and unstructured that can be analysed to bring value. The typical definitions (e.g., by NIST \cite{nist_big_data_public_working_group_definitions_and_taxonomies_subgroup_nist_2018} or Gartner \cite{gartner_it_glossary_what_2018}) refer to big data by a number of \textit{V}-properties, such as 
volume, velocity, and variety. 
Today, big data has become capital, with enterprises improving substantially their operations and customer relations, and the academia developing and enhancing research (e.g., in climate \cite{overpeck_climate_2011} or biology research \cite{marx_biology:_2013}).
In addition, the huge amount, generation speed, and diversity of data require special architectures for storage and processing (e.g., MapReduce \cite{dean_mapreduce:_2008} or Apache Hive\footnote{\url{https://hive.apache.org/}}). 

While the usefulness of processing big data is mainly unquestioned, it also comes with high privacy risks when operating on personal data. This is mainly due to two aspects of big data analysis. First, the larger the amount of data the higher the probability of re-identifying individuals even in datasets which seem not to have personal linking information. Second, big data analysis is able to infer from ``harmless'' personal data new information that is much more critical and was not intended to be revealed by the affected person. A famous example is the analysis of shopping patterns for creating customized (targeted) advertisements by a department store, where the algorithms correctly inferred that a teenage girl was pregnant \cite{hill_how_2012}. There are areas where privacy threats may become even more critical, such as in medical treatment or research \cite{mostert_big_2016}.

In order to protect individuals and their data a number of technical means and regulations for privacy-preserving data processing have been initiated and developed. However, implementing these methods in a data processing system obviously requires additional effort during the design phase, and in many cases such methods influence the performance of the system. As a result, in the past, enterprises and other organizations were not always willing to make this effort, but this tend to change due to the pressure applied from new privacy laws and regulations.

This paper describes privacy issues in big data analysis and elaborates on two case studies (government-funded projects\footnote{\label{fn:oslo_analytics}\url{https://www.mn.uio.no/ifi/english/research/projects/oslo-analytics/}}\footnote{\label{fn:swan}\url{https://www.ntnu.edu/iik/swan}}) in order to elucidate how legal privacy requirements can be met in research projects working on big data and highly sensitive personal information. Finally, it discusses resulted impacts on the processing of data and the results due to the employed privacy-preserving techniques.


The paper is organized as follows: Section \ref{sec:foundations} presents the current state of legal and technical aspects related to processing personal information. Section \ref{sec:use_cases} presents and analyzes two research projects operating on large datasets containing personal information. In Section \ref{sec:privacy_analysis}, we discuss the influence the utilized privacy-preserving techniques had on the data processing and results of the projects. Finally, Section \ref{sec:conclusion} concludes the paper.

\section{Privacy Issues in Big Data Analytics}
\label{sec:foundations}

\subsection{Legal Regulations}
From a legal point of view, in this paper we focus on the EU \textit{General Data Protection Regulation} (GDPR) \cite{gdpr}, which came into force in May 2018. It is relevant to all organizations inside the European Union (EU), the European Economic Area (EEA) and also to organizations from other countries, if they process data of European citizens. Thus, the GDPR has effect on most major companies worldwide. 


The GDPR regulates the collection, storage, and processing of personal data. Personal data are any data that can be linked to a specific natural person. This includes not only direct personal identifiers (e.g., full name, national ID number) but also indirect identifiers like phone numbers, IP addresses, or photos with identifiable people. Data that do not include such identifiers are commonly regarded as \textit{anonymous} and are outside the scope of GDPR (Recital 26). The results of big data analysis are very often statistical findings without direct links to specific individuals. Hence, a simple method to conform to all requirements of GDPR is to process only anonymous data.
However, the definition of anonymity is not trivial. Even if directly identifiable parameters are removed from a dataset, it might be possible to \textit{re-identify} single individuals by combining the dataset with other information. This approach for de-anonymization is called \textit{background knowledge attack} \cite{machanavajjhala_l-diversity:_2006,Kifer:2011:NFL:1989323.1989345}.

A famous example of re-identification is the Netflix challenge in 2006. As part of a competition for finding more accurate movie recommendation methods, Netflix released a dataset containing movie ratings of 500,000 customers. In the dataset, any personally identifiable information (PII) was removed and only subscriber IDs (without any connection to the actual identity) and movie ratings (score, movie info, date) were published. However, researchers combined these data with other publicly available information (e.g., IMDB ratings) and were able to identify individual customers with a high probability \cite{narayanan_robust_2008}. Other well-known cases include identification of individuals from internet search terms \cite{barbaro_face_2006}, anonymized DNA \cite{bohannon_genealogy_2013} and mobility data \cite{montjoye_unique_2013}.

There are numerous formal metrics for measuring the degree of anonymity of a dataset (see next section). GDPR without giving a precise or concrete definition of anonymity considers a dataset anonymous when re-identification is only possible with high effort or unlikely means.

\begin{figure}[t]
\centering
\includegraphics[width=1\linewidth]{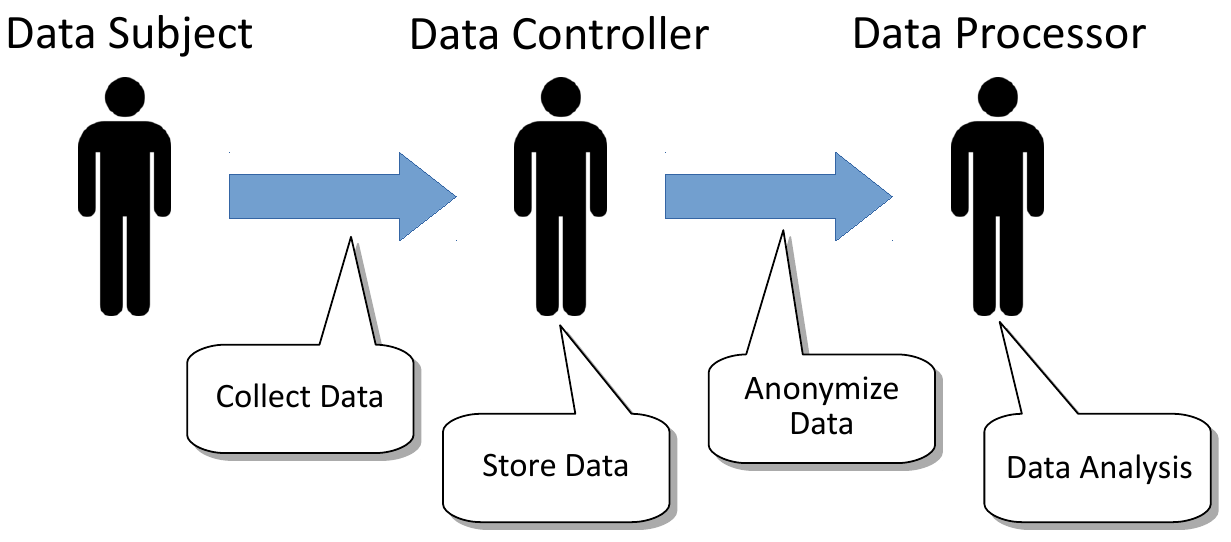}
\caption{Personal data handling process}
\label{fig:data_handling_gdpr}
\end{figure}

For processing personal data the GDPR defines a number of legal, organizational and technical requirements, and proposes different methods. The most relevant principles are described here. First of all, in most cases, processing of personal data is allowed only if the data subject has given its \textit{consent} (Article 6)\footnote{\underline{Article 6} of the GDPR regulates the lawfulness, fairness and transparency of collecting and processing personal data.}. Exceptions apply when the data processing is explicitly allowed by a law or regulation, or ensures ''vital interests of the data subject''. Additionally, the consent given must be limited to a specific purpose for data processing (Article 5)\footnote{\underline{Article 5} of the GDPR regulates principles relating to processing of personal data.}. The data controller (the entity that is responsible for collecting the data) can neither define a too generic data processing purpose nor change the purpose later arbitrarily (see Figure \ref{fig:data_handling_gdpr}). 

Another data processing principle is \textit{data minimization} (Article 5) which refers to limiting personal data collection, storage, and usage to data that are relevant, adequate, and more importantly necessary for carrying out the purpose for which the data are processed.
Worthy of noting is that \textit{pseudonymization} is explicitly mentioned as a data minimization measure. In pseudonymized data, identifiable parameters are replaced by other (randomly) generated identifiers. This usually does not have any negative impact on the data mining process and preferably should be initiated by the data controller before transferring the data to the data processor. If the data processing results require to be linked this can be achieved by the data controller, as it holds the mapping (also known as \textit{a pseudo-lookup table}) pseudonyms to the identifiable parameters. In addition to the anonymized data, storage at the data controller must conform to the GDPR by employing techniques that protect the data in rest (e.g, encryption and tight access control).
Furthermore, the GDPR requires ''\textit{appropriate technical and organizational measures to ensure a level of security appropriate to the risk}'' (Article 32)\footnote{\underline{Article 32} regulates the security of data processing.}, which commonly includes the application of techniques like data encryption, access control, physical protection and (again) pseudonymization.

An extension to the data minimization principle is the \textit{storage limitation} principle, which restricts the duration of data storage to a specified (necessary) period.

In the context of data processing, it must be further taken into account that automatic decision-making processes with impact on individuals (Article 22)\footnote{\underline{Article 22} regulates automated decision-making and profiling.} as well as processing of extremely sensitive data, such as biometric data (Article 9)\footnote{\underline{Article 9} regulates the processing of special categories of personal data.}, requires ``\textit{explicit}'' consent from the data subject.

The terms big data or data analysis are not addressed by the GDPR directly. However, it is clear from the description above that big data and the GDPR are not always compatible \cite{zarsky_incompatible:_2016}. For example, big data mining relies on the analysis of large amounts of data, which often contradicts the principle of data minimization. In addition, in data analysis very often new hypotheses for testing are introduced after the data were collected. However, the data subjects from which the data were collected have given consent initially for a different purpose. Thus, from a legal perspective data processing should be done---if possible---on anonymized data, otherwise great care must be taken that the GDPR is respected. This for example might require a \textit{data protection impact assessment} (DPIA); a privacy-related impact assessment whose objective is to identify and analyse how data privacy might be affected by certain actions or activities (Article 35)\footnote{\underline{Article 35} regulates the data protection impact assessment (DPIA).}.

The fines that can be levied as penalties for non-compliance are based on the specific articles of the Regulation that the organization has breached. The GDPR also gives individuals the right to compensation for material and/or non-material damage resulting from an infringement of the Regulation. Specifically, data controllers and processors face administrative fines of

\begin{itemize}
\item up to 10 million Euros or 2\% of annual global turnover (whichever amount is higher) for infringements of articles: 8 (conditions for children’s consent), 11 (processing that does not require identification), 25-39 (general obligations of processors and controllers), 42 (certification), and 43 (certification bodies).
\item up to 20 million Euros or 4\% of annual global turnover (whichever amount is higher) for infringements of articles: 5 (data processing principles), 6 (lawful bases for processing), 7 (conditions for consent), 9 (processing of special categories of data), 12-22 (data subjects’ rights), and 44-49 (data transfers to third countries).

\end{itemize}

\subsection{Technical Aspects}
The legal requirements presented in the previous section must be implemented by technical means. This section presents some methods for privacy-preserving data mining. Well-known early approaches in this area are the works of Agrawal and Skrikant \cite{agrawal_privacy-preserving_2000}, and Lindell and Pinkas \cite{lindell_privacy_2002}. In the first one, data were anonymized by distortion, and a special decision tree classification analysis was performed on the anonymized data. In the second one, the data were split over two separate databases (which can be seen as a type of pseudonymization) and a special multi-party computation algorithm was developed for analysing the dataset. This shows the typical parts of privacy-preserving data analysis: anonymization (as effective as possible; at least pseudonymization) and potentially mining algorithms adapted to this kind of modified data.

To support the anonymization process the attributes in a dataset are usually divided into four distinct categories \cite{malik_privacy_2013}:

\begin{itemize}
	\item \textit{Explicit identifiers}: attributes that each directly link to a single individual, like social security number or email address.
	\item \textit{Quasi-identifiers}: attributes that do not directly link a person, but can re-identify an individual when the values of multiple attributes are combined. Examples are: date of birth, ZIP code or profession.
	\item \textit{Sensitive information}: attributes containing information the data subject does not want to be revealed or at least not be linked with its person. Examples might be: diseases, financial situation, sexual orientation, current position.
	\item \textit{Non-sensitive information}: attributes that do not fall in any of the aforementioned categories (e.g., weather data).
\end{itemize}
Unfortunately, this categorization is not always obvious: a (rare) disease might also identify a person. Also inside the categories there are large differences: the place of residence as a quasi-identifier can refer to millions of people if referring to a large city, but can also point to only a handful of individuals for small villages. To quantify the rate of anonymity and thereby the threat regarding re-identification, a number of anonymity models exist. The most common approaches are \textit{$k$-anonymity} \cite{sweeney_k-anonymity:_2002}, \textit{$l$-diversity} \cite{machanavajjhala_l-diversity:_2006}, \textit{$t$-closeness} \cite{li_t-closeness:_2007} and \textit{differential privacy} \cite{dwork_differential_2008}.

It is a fact that most of the times sensitive information is of high value for data mining, but also for adversaries. If the linkage between explicit identifiers and sensitive information is the goal of an analysis, obviously anonymization is not possible and the GDPR must be regarded. In this case, at least pseudonymization should be applied. Very often, however, data mining is looking for connections between quasi-identifiers and sensitive attributes, which allows anonymization of data. Common anonymization methods are \cite{xu_information_2014}:
\begin{itemize}
	\item \textit{Suppression}: removing the values of an attribute completely or replacing them with a dummy value (typically an asterisk ``*''). This operation is usually performed on explicit identifiers.
	\item \textit{Generalization}: replacing values with more general or more abstract values inside the attribute taxonomy, for example, date of birth $\rightarrow$ age (in years); age (in years)  $\rightarrow$ a range of years; ZIP code  $\rightarrow$ first two digits of the ZIP code. This operation is usually performed on quasi-identifiers.
	\item \textit{Permutation}: partitioning the data into groups and shuffling the sensitive values within each group. As a consequence, the relationship between quasi-identifiers and sensitive data is eliminated.
	\item \textit{Perturbation}: replacing values in a way that linkage to the original data is removed, but keeping the statistical properties similar. A typical method for perturbation is adding noise \cite{narayanan_robust_2008}.
\end{itemize}
The models and anonymization techniques presented are not just of academic interest, but are used in practical privacy guidelines (e.g., the Norwegian data protection authority) \cite{norwegian_data_protection_authority_anonymisation_2017}. The aforementioned anonymization operations obviously cause a loss of information and reduce the \textit{utility} of the data \cite{yu_big_2016}. Using the metrics for utility and anonymity (see above), one can evaluate different anonymization approaches and find the trade-off between privacy and utility.

Data mining on anonymized datasets sometimes requires specially adapted mining algorithms. Typical examples of classification and clustering algorithms for which privacy-preserving versions exist are decision trees, Bayesian classification, support vector machines (SVM) and secure multi-party computation \cite{xu_information_2014}.

\section{Case studies: GDPR and Real-life Research Projects}
\label{sec:use_cases}

\subsection{Oslo Analytics}
\label{sec:oa}
The \textit{Operable Subjective Logic Analysis Technology for Intelligence in Cybersecurity}\footref{fn:oslo_analytics}
is a research project (project number: 247648) funded under the ICT and Digital Innovation program of the Research Council of Norway for the University of Oslo for the period of 2016 -- 2019. Oslo Analytics develops advanced analytical methods based on big data analysis, machine learning and subjective logic \cite{josang_trust_2006} to gain a deep situational awareness and understanding of security incidents. The project is organized in collaboration with national and international institutions, organizations and security vendors, such as mnemonic, the Norwegian Computing Center (NR), the Norwegian National Security Authority (NSM), The Defence Intelligence College, the US Army Research Labs, and the Technische Universit\"at Darmstadt.

Oslo Analytics needs to conform to the GDPR and the Personal Data Act of 2000 (managed by the Norwegian Data Protection Authority - \textit{Datatilsynet}). The Norwegian Centre for Research Data (NSD) is responsible for implementing the statutory data privacy requirements in the research community, and thus requires notification from every research project processing personal data that are not fully anonymized. Fully anonymous data are information that cannot in any way identify an individual either directly through name and national identity number, or indirectly through background variables, a name list, scrambling key, encryption formula or code.

\subsubsection{Handling Sysmon Data - End Point Security}
Data of particular research importance for Oslo Analytics are Sysmon logs. Sysmon is a Windows system service and device driver that monitors and logs system activity of Windows workstations, servers and domain controllers. Sysmon provides some of the most effective events needed to trace attacker activity and increase host visibility. For example, Sysmon event class "Process Create" with ID 1 can detect initial infection and malware child processes by capturing hashes. Sysmon event class "File Creation Time Changed" with ID 2 can detect anti-forensic activities, such as changes in the file creation time of a backdoor to make it look like it was installed with the operating system. Sysmon event class "Network Connection" with ID 3 can be used to identify network activity, such as connections to command and control servers (C\&C) or even download encryption keys from ransomware servers. Research on Sysmon aims to reduce the cumbersome process of investigative analyses (threat hunting with NoSQL database systems or graph databases) by providing new complementary means based on Artificial Intelligence (e.g., ontologies \cite{mavroeidis2018data}) and specifically machine learning. 

Like many other datasets, Sysmon contains multiple privacy-sensitive identifiers (Windows account usernames, computer names, static internal IPs) and user-behaviour (running processes, internet activity) that Oslo Analytics has to deal with prior processing. For example, sensitive fields in events with ID 1 include \textit{computer name}, \textit{command line}, \textit{current directory}, \textit{user}, \textit{parent image}, \textit{parent command line}. All the aforementioned fields can reveal the identity of the user either directly or indirectly. In addition, complete removal of the aforementioned information (fields) would disallow researching and experimenting with technologies such as natural language processing narrowing down our options to more simplistic and less effective approaches.

A fallacy identified in the very early stage was the hashing of computer names in the dataset to keep the mapping between parent and child processes, as well as the time-sequenced activity of computers. This approach (hashing computer names) could allow re-identification of the original computer names and consequently the users operating the computers and their activity by re-hashing the computer names found in new Sysmon data \cite{marx_hashing_2018}. Thus, for keeping the computer activity linkability in the dataset we generated unique integer identifiers that replaced the computer names, without keeping any mapping between them.


\subsubsection{Data Storage and Accessibility}
The data are stored on a secure server with access restricted to authorized researchers working on Oslo Analytics under a very tight access control list adopting the principle of least privilege. Processing of the data can only occur on the server. Access to the secure server is only allowed from inside the organizational network and this is restricted to specific computers filtered by their MAC addresses, their internal static IP, and user account. In addition, a firewall has been configured to allow only incoming connections to the server on port 22 (SSH). Any other network activity is denied and consequently dropped. In this respect, the network restrictions disallowed us to personally install any extra programming libraries needed for processing the data after setting up the server. Thus, we had to inform the security team that is responsible for the security of the server and the data stored. Finally, the user accounts for processing the data on the secure server are only valid for the duration of the project (account expiration), meaning that the accounts will be disabled on a specific date. The same principle applies to the Sysmon data which restricts the duration of the data storage to the active period of the project.

\subsubsection{Trade-off between Security, Reproducibility and Dataset Availability}

Reproducibility provides transparency to data analyses and allows the transfer of knowledge to others who could learn from your data and methods. Reproducible research demands that data analyses and scientific claims are published with their raw data and software code so others interested may verify the findings and build upon them.


Even anonymized Sysmon datasets can be a great source of information for any malicious actor interested to harm an organization. It can be used to identify vulnerable and unpatched applications running on workstations and servers, to determine the version of Windows operating systems running in the organizational environment, to process network activity, to identify file names, etc. Network activity even anonymized can be used with various success for phishing attacks since insights for the most visited domains can be obtained. In case of re-identification network activity can be used successfully for crafting more targeted phishing attacks. For the aforementioned reasons Oslo Analytics could not make publicly available the anonymized Sysmon dataset used in the research.

\subsection{SWAN}
\label{sec:swan}
The \textit{Secure Access Control over Wide Area Network (SWAN)}\footref{fn:swan} is a research project funded by the Research Council of Norway (Grant number: IKTPLUSS 248030/070) for the Norwegian University of Science and Technology (NTNU) for the period of 2015 -- 2019. SWAN is composed of the following six partners; NTNU as the coordinator of the project (Norway), the University of Oslo (Norway), the IDIAP Research Institute (Switzerland), the Association of German Banks (Germany), IDEMIA (France), and Zwipe (Norway).

The SWAN project develops authentication technologies for banking and other services by using biometric identifiers. Biometric references (also known as templates) are stored, controlled and verified locally (e.g., smartphones) based on a pre-shared secret, which can be used to seal and authenticate transaction data. This overcomes the need of centralized storage for the biometric data \cite{kindt2013privacy}. The SWAN biometric authentication solutions are designed to be privacy compliant and align with existing and emerging biometric standards.

SWAN needs to conform to the GDPR and the Norwegian Personal Data Act in the same way as Oslo Analytics. The creation of the biometric dataset has been permitted by the Data Protection Official for Research (NSD).  

\subsubsection{Data Collection, Processing, and Storage}
Clause 1 of Article 9 of GDPR states that biometric data are to be considered a \textit{''special category of personal data''} and are prohibited from being used for identifying individuals, unless the data subjects have given explicit consent.
In the first phase SWAN had to collect biometric data from 200 people (data subjects).

\textit{Biometric Data Collection Consent:} The SWAN team created a dedicated Biometric Information Privacy Policy to comply with the Privacy Act and lawsuits that were into force in 2015 (before GDPR came into effect).
The policy includes the following sections and clauses:
\begin{itemize}
    \item Definition of ''biometric identifier'' and ''biometric information''
    \item Consent
    \item Disclosure
    \item Storage
    \item Retention Schedule
\end{itemize}
The data subjects were asked to aid in the construction of a biometric dataset which will be used for research purposes related to biometrics recognition and presentation attack detection (PAD) for face, voice, eye and fingerprint biometric characteristics. Prior to handing over any biometric data all participants (data subjects) signed a consent form and were informed both orally and written about the purpose of the collection. 

It is worth mentioning that in the consent form it is clearly stated that in case the data will be used for new purposes the data subjects will be asked to assess and sign a new consent form.  

The creation of the SWAN database is in accordance with the aforementioned GDPR data processing techniques, such as pseudonymization, meaning that the personal data could allow the re-identification of individuals when required. There are three main reasons for using pseudonymization measures for constructing this biometric database. First, the pseudo ID can be used to facilitate the destruction of data in the case of participation withdrawal from the project. (Article 7)\footnote{\underline{Article 7} of the GDPR regulates the conditions for consent.} clause 3 of the GDPR specifies that \textit{''consent can be withdrawn at any time''}. In such cases, all personal data including biometric identifiers related to the data subject are permanently deleted. Secondly, if the database holding pseudonymous data together with biometric characteristics is compromised, the attackers would not have the ability to look up the pseudo value and identify the data subjects. Thirdly, pseudonymization enables big data analysis without access to the raw data that contains sensitive personal information (biometric characteristics in this case). Since each data controller (project partner) have its ''own'' unique key, data cannot easily be linked among different data controllers, thus, further reducing the risk of re-identification, while affording the sharing of dedicated pseudonymous datasets (dedicated to processing for a specific purpose by an identified data controller).

\begin{itemize}
    \item \textit{Data collection:} is performed from all partners participating in the project. All the data subjects have used a purpose-built application on a smartphone to capture images and video recordings of their face, eyes and fingers, and audio recordings of their voice. In addition, the participant's name, email, gender, and age will be stored along with a pseudo ID, linking to the biometric data.

    It is worthwhile to mention that during the biometric data collection (voice data collection phase) the participants had to say four sentences: 1) "My name is \underline{$A$}, and I live in \underline{$B$}", 2) "My bank account number is \underline{$C$}", 3) "The limit of my card is 5000 Euros", and 4) "My PIN code is 9, 8, 7, 6, 5, 4, 3, 2, 1, 0". Where \underline{$A$} indicates a fictitious name, \underline{$B$} indicates a fictitious address, and \underline{$C$} indicates a fictitious bank account number. Name, address, and the bank account number are considered to be PII (personal identifying information) according to the GDPR. Therefore, these fields contain pseudo-identifiers (fiction data) generated by a random data generator in order to comply with the GDPR pseudonymization methods.
    \item \textit{Data storage:} the collected biometric data (pseudonymized) are shared among the SWAN partners, stored securely, and raw data are only accessible to researchers participating in the project from the aforementioned project partners.  
    \item \textit{Data processing:} all the partners working on the project are able to process the SWAN database based on their needs for specific work packages defined in the project description. 
\end{itemize}
All project partners that collected biometric data are responsible for their collected sub-dataset (data controllers), in terms of processing and storing the biometric data. Additionally, NTNU as a project leader serves as the main data controller.  



Biometric data may also be shared through the BEAT platform\footnote{\url{https://www.beat-eu.org/platform/}}, a research platform facilitating open research without compromising security and privacy of data as no access to raw data is given. The SWAN project is scheduled for completion during the 4th quarter of 2019, however in the SWAN project's consent form is specified that the collected data may be stored after the completion of the project for an additional maximum period. This would require the data subjects to sign a new consent form. 

In compliance with the GDPR, personal data must be kept \textit{''no longer than is necessary for the purposes for which the personal data are processed''} (Article 5 clause 1(e)). However, Article 5 also provides an exception to this rule allowing extensive data retention insofar as the personal data will be processed solely for archiving purposes in the public interest, scientific or historical research purposes or statistical purposes subject to implementation of the appropriate technical and organizational measures required by the GDPR in order to safeguard (Article 89)\footnote{\underline{Article 89} sets out safeguards and derogations relating to processing for archiving purposes in the public interest, scientific or historical research purposes or statistical purposes.} the rights and freedoms of individuals.

\subsubsection{Privacy-Preserving Biometrics}
Since biometric data are highly sensitive and cannot be easily changed, there is a need for privacy-preserving solutions to avoid misuse, loss or theft. 
The SWAN project applies biometric template protection methods to secure biometric references stored locally on smartphone devices. This can prevent misuse of biometric data in case of data theft. Biometric template protection can also prevent linking a user's biometric characteristics between different databases (cross matching), thereby preserving the privacy of the user. In addition, the SWAN project applies novel \textit{cancellable biometric} techniques (biometric template protection using Bloom Filters \cite{stokkenes2016multi}). Cancellable biometrics provide an intentional, systematic and repeatable distortion of biometric features in order to protect user's sensitive data. For example, if a ''cancellable'' characteristic is stolen, the distortions provided are modified and remapped to a new template which will replace the one that has been compromised.

\section{Lessons Learned}
\label{sec:privacy_analysis}
We have described two rather different projects dealing with sensitive large datasets. The SWAN project processes biometric data. This kind of data cannot be anonymized as the biometric samples are personally identifiable information and, thus, the GDPR applies. 
The project applied the following privacy protection methods:
\begin{itemize}
    \item \textit{Explicit consent}: the participants were informed in detail about the processing steps executed on their biometric data and had to explicitly agree on this.
    \item \textit{Security of the processing system}: the data are protected from unauthorized access using access control and encryption technologies.
    \item \textit{Pseudonymization}: to impede re-identification the mapping between biometric data and the owner (directly identifying information) is replaced by a pseudonym.
    \item \textit{Processing biometric templates}: SWAN applies different techniques to protect the biometric templates, such as cancellable biometrics which allow the revocation of a compromised biometric template. This technique does not remove the link to the data subject completely, but makes re-identification much harder. 
    \item \textit{Limited storage duration}: SWAN defines a maximum period for storing the biometric data that extends beyond the duration of the project in case additional research needed to be conducted. The dataset will be stored only for this period (and maximum up to the predefined period stated in the signed consent form) and then will be deleted. In addition, if the research deviates even slightly from the original purpose this would require the data subjects to assess and sign a new consent form.
\end{itemize}
The aforementioned methods allow SWAN to comply with the GDPR, and consequently utilize the collected data for research purposes. It is worth mentioning that under GDPR projects that collect and/or process biometric data should carry out a \textit{data protection impact assessment (DPIA)}.


The Oslo Analytics project does not operate on data that were collected explicitly for research purposes, but diversified data used in security operations. Thus, for the data processed the subjects have only given consent for purposes which are required for monitoring and protecting the network from major disruptions and attacks. As explained before, this is a typical situation in projects dealing with big data. For conforming to the legal requirements Oslo Analytics applied the following privacy protection methods:
\begin{itemize}
    \item{\textit{Anonymization}}: all data fields which allow easy re-identification of the subject have been removed (\textit{suppression}) or abstracted (\textit{generalization}).
    This does not only apply to directly identifying data like usernames but also to fields like internal IP addresses.
    \item{\textit{Security of the processing system}}: like in the SWAN project, access to the processing system was strictly controlled and restricted.
\end{itemize}
Like mentioned before, re-identification might be possible in large anonymized datasets. Nonetheless, the datasets used for processing should fulfill the GDPR with its rather weak (not a strong formal definition) anonymization requirement.

Like in the SWAN project, projects that operate on data which are sensitive, processed on a large scale, and fall under the special categories referred to in Article 9, clause 1 of the GDPR should carry out a data protection impact assessment (DPIA). In addition, it is the case that anonymizing methods can prohibit the use of specific technologies, such as natural language processing which would be beneficial for improving the successfulness and potentially the results of the research.

\section{Conclusion}
\label{sec:conclusion}

This paper presented the implications of data protection laws on projects dealing with big data, and by using two case studies analysed how privacy-preserving techniques can be applied. The results were quite different. In one project, for mitigating privacy concerns regarding biometric data collection and processing the participants were asked to give consent. In addition, no problems were faced during the data analysis phase. In the second project, data from an existing data source were used. Here, anonymization of many data fields was required, making the data analysis more challenging and in many cases limited. It is of great importance to remark that for projects and technologies dealing with sensitive data a data protection impact assessment should be conducted at the very early stages of the project to identify potential privacy challenges, and to adapt the analysis methods taking into consideration privacy-preserving techniques. 


\section*{Acknowledgement} 
This research was partially supported by the Research Council of Norway under the Grant No.: IKTPLUSS 247648 and 248030/O70 for Oslo Analytics and SWAN projects, respectively.

\balance 
\bibliographystyle{IEEEtran}
\bibliography{bibliography}

\end{document}